# Band engineering of phosphorene/graphene van der Waals nanoribbons toward high-efficiency thermoelectric devices

Maryam Mahdavifar [1], Farhad Khoeini*[,1], Francois M. Peeters [2,3]

[1] Department of Physics, University of Zanjan, P.O. Box 45195-313, Zanjan, Iran

[2] Centre for Quantum Metamaterials, HSE University, Moscow 101000, Russia

[3] Departamento de Física, Universidade Federal do Ceará, Fortaleza, Ceará, 60455-760, Brazil

**Abstract**

Vertical integration of dissimilar layered materials in a so-called van der Waals (vdW) heterostructure (HS) has emerged as a useful tool to engineer band alignments and interfaces. In this paper, we investigate thermoelectric currents in a phosphorene/graphene vdW nanoribbon consisting of an armchair graphene nanoribbon (AGNR) stacked on an armchair phosphorene nanoribbon (APNR). We focus on the currents driven by a temperature difference between the leads in a two-probe junction. In contrast to pristine AGNRs and APNRs, such an armchair-edged HS can provide several nano-amperes of the current at room temperature, without any external field. External electric fields modify the electronic band structure and are able to induce a type-I to type-II band alignment transition and a direct-to-indirect band gap transition. Biasing the APNR/AGNR by an external electric field is found to strongly increase the thermally induced current and also control the direction of current flow at moderate temperatures. These results are important for potential applications of the APNR/AGNR vdW HS in flexible electronics and thermoelectric devices.

**Introduction**

Two-dimensional (2D) nanomaterials have been recognized as potential candidates for next-generation electronics because of their exotic electrical, mechanical, optical, and thermal properties, which are not revealed in their bulk counterparts [1-6]. Among the family of 2D layered structures, graphene has emerged as the most appealing material in the last two decades. Graphene is defined as a single layer of $sp^2$-hybridized carbon atoms packed into a honeycomb structure, which has been demonstrated to host intriguing electronic and quantum transport properties, such as massless Dirac fermions, high charge carrier mobility, and half-integer quantum Hall effect [7-10]. However, the gapless nature of this key material reduces its utility for practical applications [10, 11]. Hence, many efforts have been dedicated to create a small electronic band-gap in graphene, e.g., involving size effect [12-15].

The isolation of graphene by mechanical exfoliation of graphite in 2004 [1] brought forth a significant advancement of thin-layer processing methods, giving rise to a renewed research interest in graphite-like layered materials such as black phosphorus (BP) [16, 17]. BP, the most thermodynamically stable allotrope of phosphorus [18-20], is made of single-atom-thick sheets, which are held on top of each other by weak van der Waals forces [21, 22]. Due to the weak forces of attraction between layers, few-layer BP, known as phosphorene, can be easily cleaved from bulk crystals [4]. Due to the $sp^3$ hybridization of phosphorus atoms, it has a hexagonal arrangement with a puckered structure, resulting in in-plane anisotropic spectrum [23, 24]. As a semiconductor, it exhibits a nearly direct band gap, which is strongly thickness-dependent, increasing from ~0.3 eV to ~2 eV with decreasing thickness [25-27]. Moreover, it shows interesting properties, including a high carrier mobility of ~1000 $cm^2V^{-1}s^{-1}$, a moderate on/off current ratio of ~$10^4$, and negative reflection [4, 28, 29]. With its lower effective masses of carriers than transition metal dichalcogenides [25, 30], phosphorene is more suitable for engineering devices. Despite these extraordinary properties, a main issue hindering phosphorene application in electronic devices is its poor air stability [31, 32].

Integrating dissimilar 2D materials in a so-called vdW heterostructure is a useful tool to engineer band alignments [33-35]. Compared to single-material systems, HSs seem to be more promising for functional material applications due to the possibility of electronic band engineering. Along with the rapid development of vdW HSs, several different types of phosphorene-based HSs have been recently studied in detail and utilized in electronic devices, e.g., vdW phosphorene/graphene Schottky junction [36], phosphorene/MoS$_2$ junction as an effective solar cell material [37], and phosphorene-based field-effect transistors [38, 39]. Interestingly, an HS constructed with a proper layer-by-layer assembly not only may serve to overcome the intrinsic shortcomings of the materials but also provide novel functionalities such that the relevant properties of each component material can be effectively maintained. For instance, to protect phosphorene from structural and chemical degradation, a graphene layer or a hexagonal boron nitride layer has been proposed to be stacked on top of it preserving the distinctive features of phosphorene [40]. The HSs have also received significant attention in designing high-performance thermoelectric devices. Zeng et al. predicted a significant enhancement in the thermally induced current of H-terminated zigzag graphene ribbon (ZGNR) contacted to O-terminated ZGNR, compared to a perfect H-terminated ZGNR [41]. Hu et al. reported an enhancement in the *ZT* value of black phosphorus/blue phosphorus vdW HS compared to pristine layers [42].

In this paper, we investigate current driven by a temperature difference in APNR stacked on AGNR by solving the electronic transmission in the nonequilibrium Green's function formalism. We use the tight-binding (TB) model to describe the electronic structure of the system. It was reported that APNRs [43] and AGNRs [44] show excellent thermoelectric performance by gating or doping. A recent study has shown a high thermoelectric performance in a phosphorene nanoring contacted to APNR with modulation of the chemical potential [45]. Unlike pure APNR and AGNR, APNR/AGNR vdW HS is found to offer several nano-amperes of thermally induced current at moderate temperatures, without the presence of any field. We also investigate the effect of external electric fields on the thermally induced current, which revealed a significant enhancement with electric field.

**Model and method**

2D phosphorene/graphene vdW HS consists of a monolayer graphene lying on a monolayer phosphorene. Since we are interested in the minimal lattice mismatch between the layers, we suppose that the unit cell of the HS is composed of a 1×4 graphene supercell stacked on top of a 1×3 monolayer phosphorene supercell, as shown in figure 1(a). The remaining mismatch should be compensated by strain. Due to the weak dependence of the electronic spectrum of graphene on strain [46] (unlike phosphorene [23]), it is applied only to the graphene layer. Considering the equilibrium geometry obtained from DFT calculations reported in [36], the interlayer distance, $\Delta_{PG}$, is assumed to be 3.45 Å.

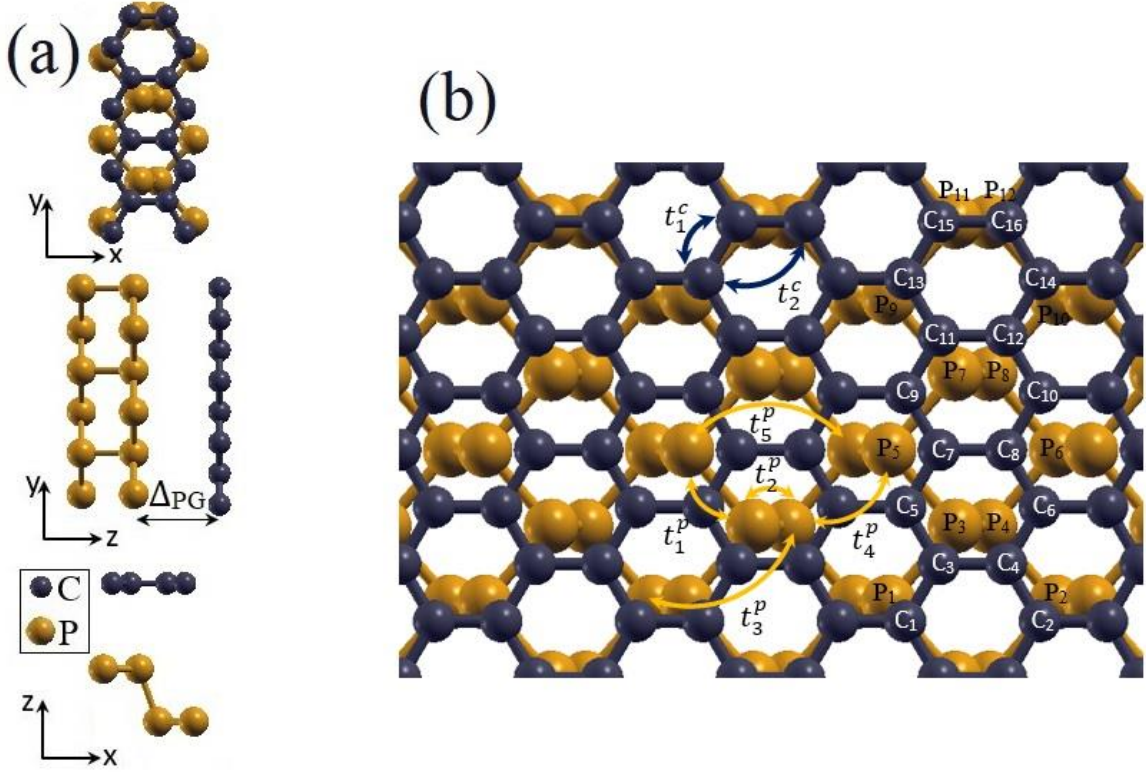

Figure 1. (a) Different views of the supercell of phosphorene/graphene vdW HS consisting of a 1×4 monolayer graphene supercell stacked on top of a 1×3 monolayer phosphorene supercell. (b) Schematic of 2D HS. Black lattice is graphene and the orange one is phosphorene. The atomic sites are labeled per supercell, and the different hopping parameters are indicated.

We denote an armchair-edged phosphorene/graphene vdW HS by $N_P$APNR/$N_G$AGNR, where $N_P$ and $N_G$ are defined as the number of rows of atoms parallel to the x-axis (figure 1(a)) for APNR and AGNR, respectively. For instance, the armchair-edged nanoribbon with the same width of the supercell shown in figure 1(a) (the periodicity is along the x-axis) has a width index of $N_P = 6$ and $N_G = 8$, denoted by 6APNR/8AGNR, and resulting in a width (W) of ~8.6 Å.

The low-energy Hamiltonian is described within a simplified TB model, including five hopping parameters for the phosphorene layer [47], first and second- neighbor hopping parameters for the graphene layer [48], and a coupling parameter between two layers specified within the method introduced in [49]. We emphasize that the model is valid only at low energies. The explicit representation is given in the appendix.

In this work, using the nonequilibrium Green's function method, we study the evolution of the thermally induced current through a two-probe junction depending on the width as well as under biasing, which is implemented by external electric fields. We set up the two-probe junction by partitioning APNR/AGNR into a central region and two semi-infinite terminations as source and drain leads. The formalism is as follows.

The Green's function of the junction is defined as [50-51]

$$\boldsymbol{G}(E) = [(E + i\eta)\mathbf{I} - \boldsymbol{H} - \boldsymbol{\Sigma}_L(E) - \boldsymbol{\Sigma}_R(E)]^{-1}, \tag{1}$$

where $\eta$ denotes an arbitrary infinitesimal number, $\mathbf{I}$ is the identity matrix, and $\boldsymbol{H}$ is the real space Hamiltonian matrix. Additionally, $\boldsymbol{\Sigma}_L$ ($\boldsymbol{\Sigma}_R$) is the self-energy term, which can be interpreted as the effective Hamiltonian arising from the coupling of the central region with the left (right) lead. We compute the self-energies using the iterative procedure introduced in [52-54]. The electronic transmission probability [55-57] is given by

$$T_e(E) = \text{Tr}[\boldsymbol{\Gamma}_L(E)\boldsymbol{G}(E)\boldsymbol{\Gamma}_R(E)\boldsymbol{G}^\dagger(E)], \tag{2}$$

where $\boldsymbol{\Gamma}_{L(R)}$, the broadening function of the left (right) lead, is given by

$$\boldsymbol{\Gamma}_{L(R)}(E) = i\left[\boldsymbol{\Sigma}_{L(R)} - \left(\boldsymbol{\Sigma}_{L(R)}\right)^\dagger\right]. \tag{3}$$

In addition, the density of states (DOS) at energy $E$, is described as

$$\text{DOS}(E) = -\frac{1}{\pi}\text{Im}\left(\text{Tr}(\boldsymbol{G}(E))\right). \tag{4}$$

The local density of states (LDOS) is understood as the spatial distribution of electronic states at a given energy.

The temperature difference between the leads gives rise to a current, $I$, through the contact given by

$$I(T) = 2\frac{e^2}{h}\int dE\, T_e(E)\big(f_L(E-\mu) - f_R(E-\mu)\big), \tag{5}$$

where $\mu$ and h are the chemical potential and the Plank constant, respectively. The factor 2 counts for the spin degeneracy. Besides, $f_{L(R)} = 1/(1 + \exp\left(\frac{E-\mu}{k_B T_{L(R)}}\right))$ is the Fermi distribution function related to the left (right) lead at the temperature of $T_{L(R)}$, where $k_B$ is the Boltzmann constant.

**Discussion**

**I. The electronic structure of 2D HS**

Our approach is to use the TB model for calculating the thermally induced current in the armchair-edged HS. However, in order to assess the reliability of the TB parametrization, we first use the model to calculate the electronic structure of 2D phosphorene/graphene vdW HS and compare the results with previous works based on DFT calculations. Figure 2 shows the band dispersion of the HS together with that of the pristine layers. As shown in figure 2(a), compared to the energy bands of the pristine layers, the superpositions of the HS are slightly perturbed. Moreover, a band gap increase of 0.11 eV is induced to phosphorene by forming the vdW HS. The enlarged views of the regions, indicated by the black circles in figure 2(a), representing the Dirac cone or a band anti-crossing, are shown in figures 2(b)-2(e). Figure 2(b) reveals that the Dirac point undergoes a small splitting of ~0.02 eV and becomes higher in energy by ~0.01 eV upon contact. In addition, the band-cross points open a gap by less than 0.06 eV as a result of the vdW interlayer coupling. These results are in good agreement with previous DFT calculations [36, 58].

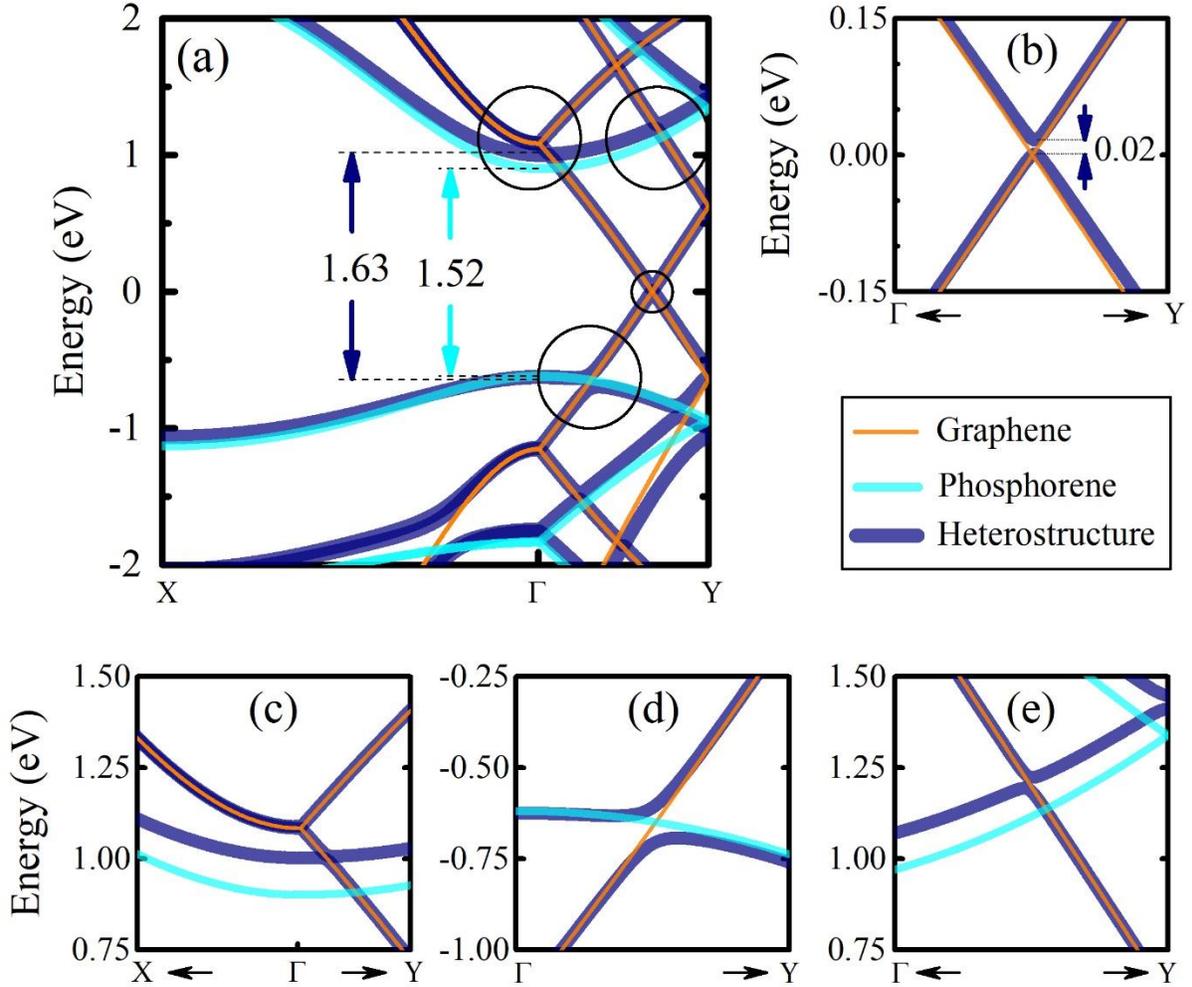

Figure 2. (a) The band dispersion of the 2D HS, graphene, and phosphorene. The enlarged views of the regions indicated by the black circles are shown in (b)-(e), representing (b) the Dirac cone, and (c)-(e) band anti-crossings.

**II. Thermally induced current in the armchair nanoribbon**

In this section, we focus on the electronic currents resulting from a temperature difference between the leads of a two-probe junction with armchair edges without any biasing. To determine the electronic properties of the armchair-edged HS, we first investigate the electronic structure and transmission coefficients of APNR/AGNR.

All APNRs exhibit semiconducting character, with an energy gap larger than 1.56 eV for W < 4 nm [59]. As a result of the symmetry restrictions, the three characteristic groups of AGNR are defined by the width rule (semiconducting for $N_{\text{Gr}} = 3n, 3n + 1$ and metallic for $N_{\text{Gr}} = 3n + 2$, where $n = 1, 2, 3, ...$) [12]. Hence, we have two possible electronic configurations for the APNR/AGNR interface depending on the width: semiconductor/semiconductor or semiconductor/metal. For instance, the band structure of 19APNR/25AGNR (a semiconductor/semiconductor junction) and its spatial electronic state distribution per supercell at conduction band minimum (CBM) and valance band maximum (VBM), projected at the component layers, are shown in figures 3(a)-3(c), respectively. Within the TB model, 19APNR shows a band gap of 1.590 eV, while 25AGNR has a narrow gap of 0.379 eV. The band edges of pristine layers are also shown by the dashed lines in figure 3(a), which reveals a type-I band alignment for 19APNR/25AGNR. In this case, CBM and VBM lie in the graphene layer, which is also seen in figures 3(b) and 3(c). Hence, it is expected that 19APNR/25AGNR inherits the electronic properties of 25AGNR at the band edge extrema. However, the coupling between the $p_z$ orbitals of APNR and the $\pi$ cloud of AGNR results in a band offset and a small decrease in the band gap (0.009 eV) upon binding.

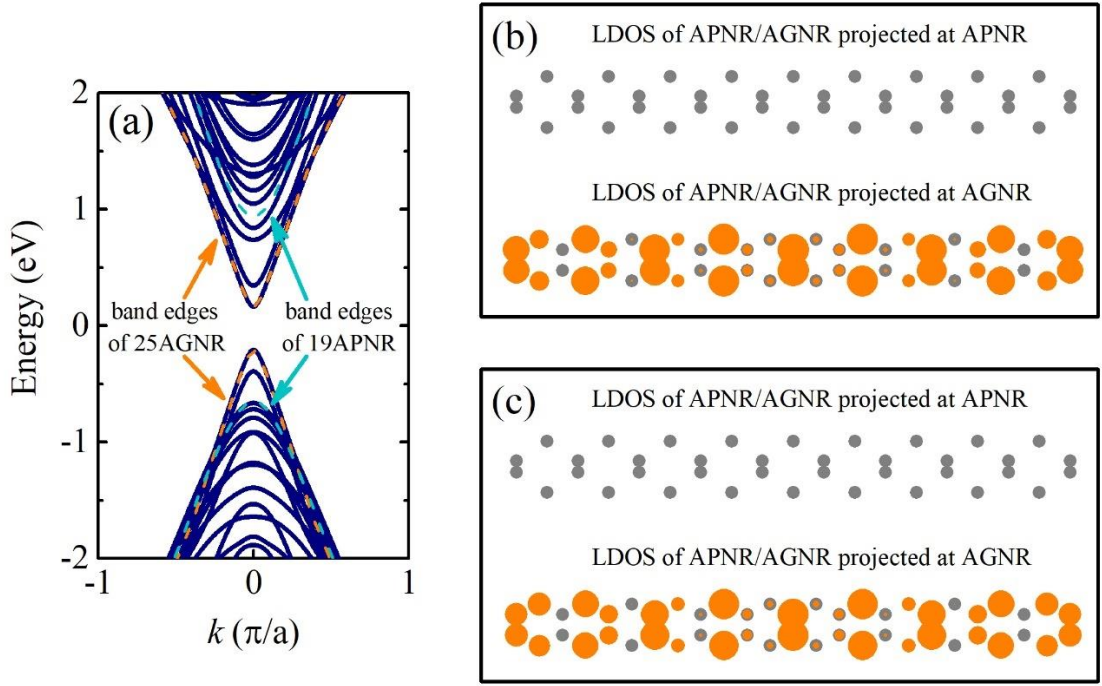

Figure 3. (a) The band dispersion of 19APNR/25AGNR, together with the band edges of 19APNR and 25AGNR, depicted by the dashed lines. LDOS of 19APNR/25AGNR at (b) CBM and (c) VBM, projected at the layers.

Considering the three characteristic groups of AGNR, we take the cases of 18APNR/24AGNR and 19APNR/25AGNR as examples of the semiconductor/semiconductor junction, and 20APNR/26AGNR as an example of the semiconductor/metal one. The electron transmission of 18APNR/24AGNR, 19APNR/25AGNR, and 20APNR/26AGNR are shown in figures 4(a)-4(c), respectively. The dashed lines plotted in figures 4(a)-4(c) exhibit the range at which the distribution function is broadened at room temperature which for illustration purpose we took $|E - E_F| < 10\, k_B T_L$ (where $E_F$ is Fermi level). From the expression of current defined by equation (5), the only charge carriers which can participate in the current have energy within the range $|E - E_F| < 10\, k_B T_L$. The charge carriers with energy higher than the Fermi energy contribute to the electron current ($I_e$), and those with energy lower than the Fermi energy contribute to the hole current ($I_h$). Considering the compensation effect (also reported in [41, 60]), the more $I_e$ overcomes $I_h$, the larger the thermally induced current, which indicates that the current should be dominated by the net transmission within the range $|E - E_F| < 10\, k_B T_L$. The thermally induced current vanishes in perfect AGNRs, arising from the symmetric transmission spectrum around the Fermi level. Although the transmission spectrum of APNRs is asymmetrical [61], it is not possible to induce a current at moderate temperature due to the presence of a relatively large gap. Indeed, the distribution function must be broadened by a significantly large temperature to generate a current in APNRs, which reduces its utility in practical applications. The nonsymmetric transmission spectrum of the heterojunction is due to the different symmetries of the wavefunctions in the component layers [62], which can offer a nonzero current.

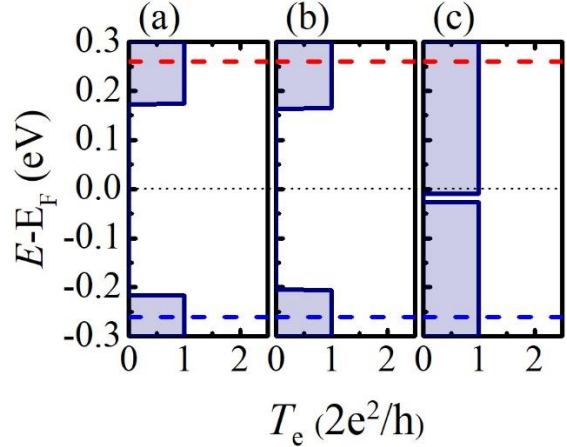

Figure 4. The electron transmission of (a) 18APNR/24AGNR, (b) 19APNR/25AGNR, and (c) 20APNR/26AGNR, as a function of energy.

Figure 5 represents the current against the width of the HS at $T_\text{L} = 300$ K for $\Delta T = 10, 20,$ and 30 K. The widths denoted by (a)-(c) correspond to figures 4(a)-4(c). According to figure 4(c), attributed to a semiconductor/metal heterojunction ($W_G = 3n + 2$), within the specified range, the transmission is close to $2\,e^2/h$ when the charge carriers have energy higher than $E_\text{F}$, whereas a small gap of about 0.02 eV is seen when the charge carriers have energy lower than $E_\text{F}$. Therefore, the electron current dominates over the hole current, which results in several nano-amperes of thermally induced current for various $\Delta T$, as shown in figure 5. Although the transmission spectra of semiconductor/semiconductor junctions are asymmetrical (see figures 4(a) and 4(b)), the thermally induced currents are close to zero or do not exceed several tenths of nano-ampere solely due to the exponential decay of the Fermi distribution function. However, since the band gap of AGNR becomes smaller as the width increases, the thermally induced current in the semiconductor/semiconductor junction gradually increases with the ribbon width. Moreover, we find that the current increases linearly with $\Delta T$. A similar linear behavior of the current-$\Delta T$ characteristics has been previously reported in layered materials [63-64].

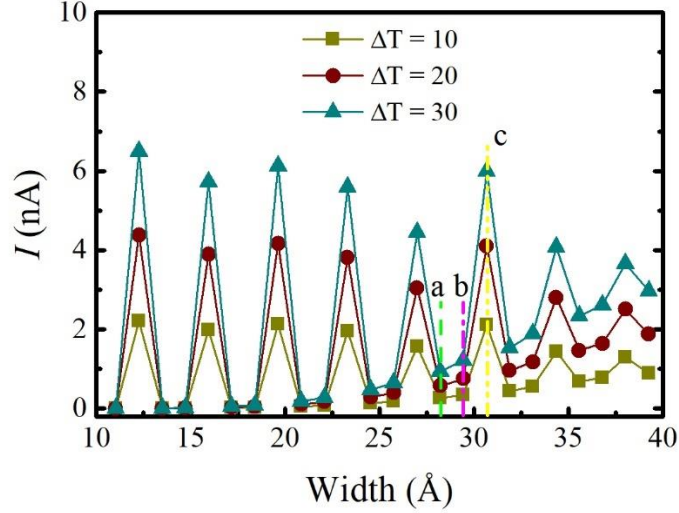

Figure 5. The thermally induced current against the width of the HS at $T_\text{L} = 300$ K.

**III. External electric field**

As we mentioned in the previous section, by breaking the electron-hole symmetry, it is possible to generate a thermally induced current. A high thermally induced current is anticipated if we enhance the electron-hole asymmetry around the Fermi level. An external electric field can be used to control the p-type and n-type doping of the HSs [65] with the possibility to engineer symmetry breaking.

Figures 6(a) and 6(b) show the electronic band dispersion of 19APNR/25AGNR in the absence and presence of a perpendicular electric field, respectively. In the presence of the electric field of 0.1 V/Å along the negative direction of the z-axis, the middle of the band gap shifts below the Fermi energy by 0.27 eV, as depicted in figure 6(b). Figures 6(c) and 6(d), respectively, correspond to figures 3(b) and 3(c) when a perpendicular electric field is applied, which indicates that the electronic states distribution at CBM remains almost unchanged. In contrast, the electronic states of both layers contribute to the electronic distribution at VBM. Indeed, upon applying an electric field along the negative direction of the z-axis, the bands of 25AGNR become lower in energy, whereas those of 19APNR become higher in energy. Further, by increasing the intensity of the electric field, the uppermost valence band of 19APNR and that of 25AGNR approach together. Up to the electric field of 0.1 V/Å, the band alignment remains type-I. The electric field of 0.1 V/Å is a critical value at which a type-I to type-II band alignment transition happens. Namely, beyond this value of the electric field, the holes (electrons) are localized in the APNR (AGNR) layer, and an efficient electron-hole separation takes place.

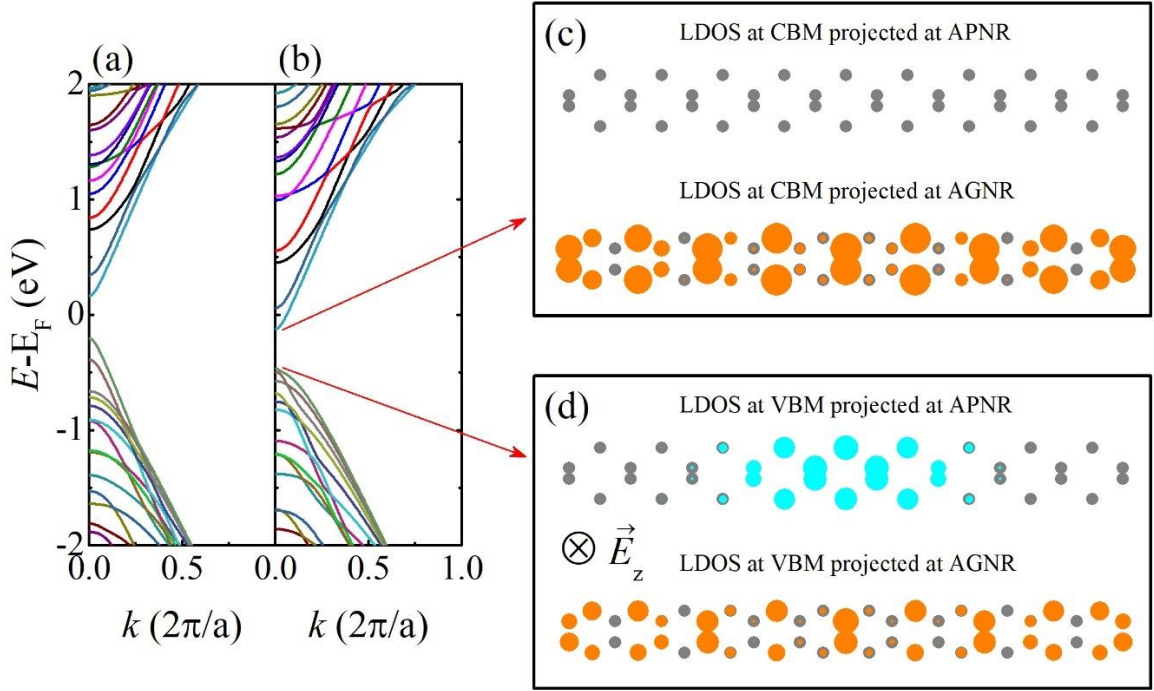

Figure 6. The band dispersion of 19APNR/25AGNR in the (a) absence and (b) presence of the perpendicular electric field of $|E_z| = 0.1$ V/Å. LDOS of 19APNR/25AGNR at (c) CBM, and (d) VBM projected at the layers. The direction of the applied perpendicular electric field is depicted in (d).

The $T_L$ dependence of the current of 19APNR/25AGNR and 20APNR/26AGNR with $\Delta T = 10$ K for different perpendicular electric fields are shown in figures 7(a) and 7(c), respectively. The dotted lines are associated with the current when $E_z$ is set to zero. One can find that the thermally induced current of 19APNR/25AGNR is relatively small. In contrast, the current of 20APNR/26AGNR reaches the maximum value of ~9 nA at $T_L = 80$ K and decreases to a few nano-amperes at higher temperature, in the absence of any applied field. The figures reveal that both the magnitude of the current and the direction of the current flow can be controlled by the perpendicular electric field. For the depicted current spectra in figure 7(a) there is a threshold temperature ($T_{tr}$), and no current is induced to the 19APNR/25AGNR at low temperatures for the given electric fields. However, variation of the $T_{tr}$ versus the applied perpendicular electric field is depicted in the inset of figure 7(a), which indicates that the threshold temperature can be reduced even to zero using the perpendicular electric field. Among the depicted curves in figure 7(a), under $E_z = -0.04$ V/Å, the magnitude of the current is relatively large (~12 nA at room temperature), and the threshold temperature is 43 K. It is evident that the current shows a higher value when $\Delta T$ is enhanced. For example, below $E_z = -0.04$ V/Å and at $T_L = 300$ K, the calculated current of 19APNR/25AGNR at $\Delta T = 30$ K is about three times higher than that at $\Delta T = 10$ K, which suggests that the current increases with $\Delta T$ almost linearly. In general, the thermally-induced current of such an armchair-edged semiconductor/semiconductor HS tends to increase as $T_L$ gets larger. Figures 7(b) and 7(d) show the electron and (the absolute value of) hole currents in 19APNR/25AGNR and 20APNR/26AGNR, respectively, as functions of $E_z$, at room temperature and $\Delta T = 10$ K. According to figure 7(b), the electron and hole currents are nearly suppressed by the electric field of $0 < E_z < 0.015$ V/Å, giving rise to a suppressed (or a tiny) net current. Indeed, there is a turning point in the $I$-$E_z$ characteristics of 19APNR/25AGNR within the range of $0 < E_z < 0.015$ V/Å, where $I_e$ equals $I_h$. When $E_z \geq 0.015$ V/Å, $I_e$ vanishes, while the magnitude of $I_h$ increases with $E_z$, and a negative current is therefore generated, which shows an increasing behavior (regardless of the sign) over $E_z$. In contrast, for $E_z \leq 0$, since $I_e$ dominates over $I_h$, a positive net current is created in the system. Moreover, $I_e$ (and of the net current) shows an increasing trend with the increase of the magnitude of the electric field along the negative direction of the z-axis. In the case of 20APNR/26AGNR, the threshold temperature is 5 K in the absence of electric field (figure 7(c)), and it vanishes by applying the electric field of $E_z = 0.002 - 0.012$ V/Å (not shown in the figure). The inset of figure 7(c) represents the critical point of the current-$T_L$ curve as a function of the perpendicular electric field in the range of $-0.005 \leq E_z \leq 0.015$ V/Å. We find that the maximal current of 20APNR/26AGNR, exceeding 20 nA (in both directions), occurs below 20 K. As shown in figure 7(d), the electron (hole) current of 20APNR/26AGNR remains almost unchanged up (down) to the electric field of 0.005 V/Å (0.007 V/Å). Up to $E_z = 0.006$ V/Å, $I_e$ highly dominates over $I_h$, and a positive net current is created in the system. In contrast, down to $E_z = 0.006$ V/Å, $I_h$ highly dominates over $I_e$, and a negative net current is created. Therefore, there is a turning point in the $I$-$E_z$ characteristic of 20APNR/26AGNR when $E_z = 0.006$ V/Å, where the electron and hole currents cancel each other. However, one can find that the maximal current created at room temperature in 20APNR/26AGNR is as low as a few nano-amperes. We conclude that for such an armchair-edged semiconductor/metal HS, a significant modulation of the thermally induced current can be obtained by utilizing a perpendicular electric field at low temperatures.

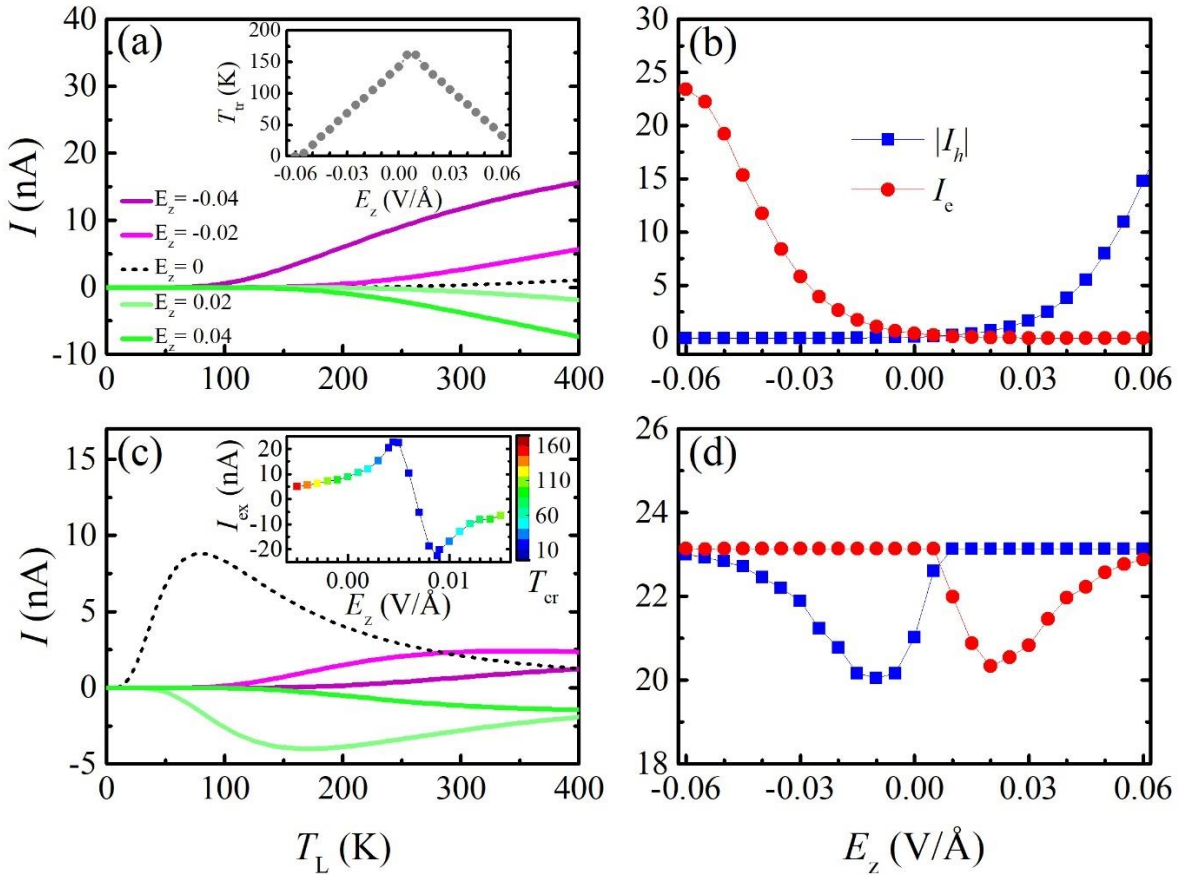

Figure 7. The thermally induced current of (a) 19APNR/25AGNR and (c) 20APNR/26AGNR versus $T_L$ for different perpendicular electric fields. The threshold temperature for the current of 19APNR/25AGNR as a function of $E_z$ is shown in the inset of (a). The critical point of the current-$T_L$ curve as a function of $E_z$ is shown in the inset of (c). The electron and (the absolute value of) the hole currents of (b) 19APNR/25AGNR and (d) 20APNR/26AGNR versus $E_z$ at room temperature for $\Delta T = 10$ K.

Now we consider the effect of the transverse electric field, along the y-axis, on the electronic properties of the ribbon. Figure 8 shows the evolution of the band edges of 19APNR/25AGNR under a transverse electric field. As we mentioned earlier, the system has a direct band gap when the applied external electric field is set to zero. Up to the transverse electric field of 0.024 V/Å, the band gap remains almost unchanged, as shown in the figure. As the electric field increases, the band gap gets smaller and gradually moves away from $k = 0$ towards $2\pi/a$ in the 1D Brillouin zone. Moreover, the figure shows that the electric field of $E_y = 0.075$ V/Å is the critical electric field at which the band gap closes and the system undergoes a semiconductor-metal phase transition. Furthermore, up to the electric field of 0.07 V/Å, the band gap remains nearly direct (not shown in the figure). When the applied electric field has a value between 0.07 V/Å and 0.075 V/Å, a direct-to-indirect band gap transition takes place.

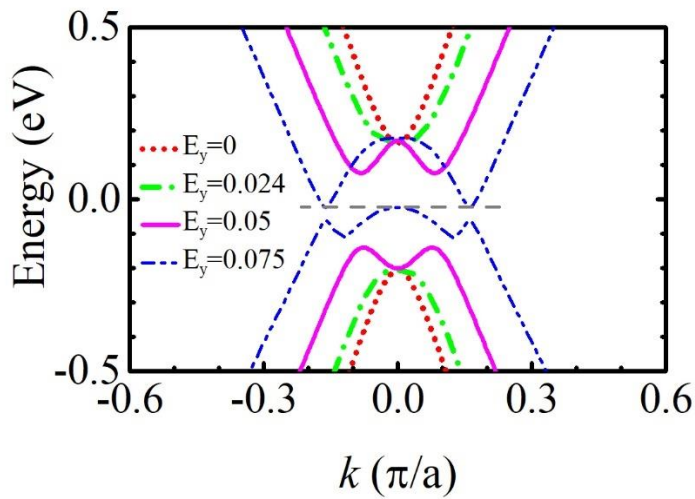

Figure 8. The evolution of the band edges of 19APNR/25AGNR under a transverse electric field.

Figure 9(a) shows the current of 19APNR/25AGNR as a function of $T_L$ with $\Delta T = 10$ K for different transverse electric fields. We mentioned earlier that the thermally induced current of 19APNR/25AGNR is relatively small without any external electric field. Figure 9(a) demonstrates that one can increase the current to several ten nano-amperes using the transverse electric field. In this case, under $E_y = 0.06$ V/Å, the maximal current of ~30 nA is obtained at the critical value of $T_L = 280$ K, and when $E_y = 0.07$ V/Å is applied, the maximal current of ~40 nA is created at the critical temperature of $T_L = 140$ K. As expected from the linear variation of the current versus $\Delta T$, when the transverse electric field is set to 0.06 V/Å, the calculated current of 19APNR/25AGNR is ~90 nA at $T_L = 280$ K and $\Delta T = 30$ K. Moreover, as shown in the inset of the figure, the threshold temperature of the ribbon gets smaller by the transverse electric field and can be reduced to zero. Among the depicted curves in figure 9(a), at room temperature, the thermally induced current is more pronounced under $E_y = 0.06$ V/Å (~30 nA), while at low temperature, the significant values of the current can be generated under $E_y = 0.07$ V/Å (~40 nA at $T_L = 140$ K). Figure 9(b) shows the electron and (the absolute value of) hole currents in 19APNR/25AGNR as functions of $E_y$ at room temperature and $\Delta T = 10$ K. According to figure 9(b), up to the electric field of 0.024 V/Å, the electron and hole currents almost remains unchanged, as a result of the unchanged band gap (see figure 8). Beyond this value of the transverse electric field, it is straightforward to see that the higher the intensity of the field, the more $I_h$ (regardless of the sign) dominates over $I_e$.

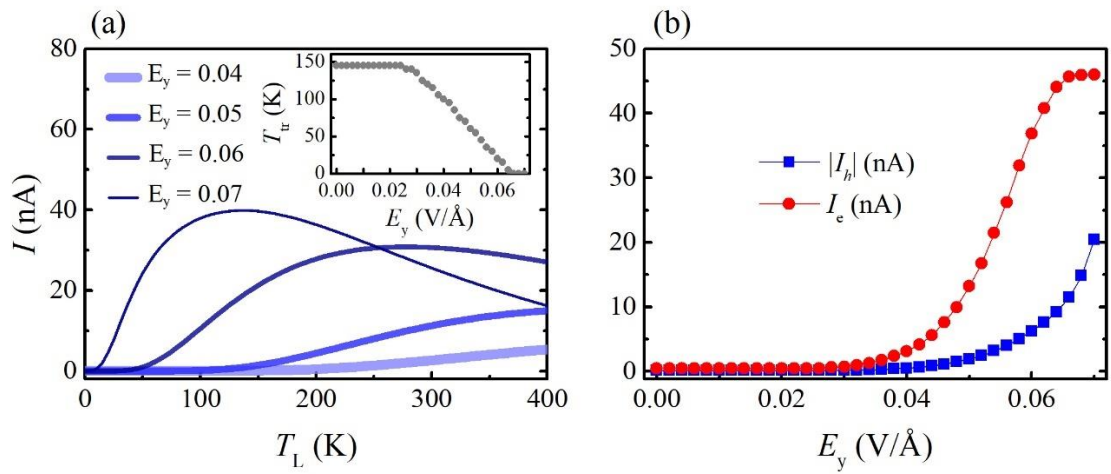

Figure 9. (a) The thermally induced current of 19APNR/25AGNR as a function of $T_L$ for different transverse electric fields. (b) The electron and (the absolute value of) the hole currents of 19APNR/25AGNR as functions of $E_y$ at room temperature and $\Delta T = 10$ K.

**Summary and conclusions**

In summary, we have investigated the thermally induced current in APNR/AGNR vdW HS using the TB model and Green's function method. To check the reliability of the TB parametrization, we used the model to calculate the electronic structure of 2D HS and found good agreement with those from DFT calculations.

We further studied the evolution of the current through a two-probe junction driven by a temperature difference between the leads. We noted that the electronic configuration of the APNR/AGNR vdW HS depends on the width of AGNR, which can be semiconductor/semiconductor or semiconductor/metal. We found that, unlike pristine AGNR and APNR, an armchair-edged HS can provide several nano-amperes of thermally induced current even without any external electric field.

We also investigated the effect of perpendicular and transverse electric fields on the electronic features of APNR/AGNR. The results indicate the possibility to induce a band alignment transition from type-I to type-II using a perpendicular electric field on APNR/AGNR. In addition, we found that the system can experience a direct-indirect band gap transition under a transverse electric field. The giant electric field-induced band modulation leads to a tunable temperature-dependent current. Biasing the APNR/AGNR by an external electric field was demonstrated to enhance the thermally induced current, even up to tens of nano-amperes.

**Acknowledgement**

Part of this work was supported by HSE University Basic Research Program.

**Competing interests**

The authors declare no competing interests.

**Appendix**

According to figure 1(a), the supercell of the HS contains 28 atoms. As shown in figure 1(b), the atomic sites per unit cell are labeled by $P_1$, $P_2$, $P_3$, …, and $P_{12}$ for phosphorus atoms and $C_1$, $C_2$, $C_3$, …, and $C_{16}$ for carbon atoms. We use a low-energy effective Hamiltonian within the TB model with one $\pi$-electron per atom. Hence, the real-space Hamiltonian is a 28×28 matrix described as

$$H = \sum_{i,j} t_{ij} c_i^\dagger c_j + eE_z \sum_i z_i c_i^\dagger c_i + eE_y \sum_i y_i c_i^\dagger c_i, \qquad (6)$$

where the summation runs over the lattice sites, and $c_i^\dagger (c_i)$ represents the creation (annihilation) operator of electrons at site $i$. The first term is the hopping energy, where $t_{ij}$ denotes the hopping parameter between two sites $i$ and $j$. The second and third terms, involving diagonal elements of the Hamiltonian matrix, describe the potentials caused by the perpendicular and transverse electric fields, respectively, where e is the electron charge, and $z_i$ ($y_i$) represents the z (y)- coordinate of site $i$. Here we provide a detailed representation of the Hamiltonian without any external potential. The hopping parameters of the system are described within a simplified TB model for the phosphorene layer, including five hopping parameters ($t_1^P = -1.220$ eV, $t_2^P = 3.665$ eV, $t_3^P = -0.205$ eV, $t_4^P = -0.105$ eV, and $t_5^P = -0.055$ eV), and first- and second- neighbor hopping parameters for the graphene layer ($t_1^C = -2.7$ eV and $t_2^C = -0.2$ eV), as shown in figure 1(b). The coupling Hamiltonian between the layers is described by the interlayer hopping parameter, $t^{PC}$, up to a distance of 4 Å.

The $K$-dependent Hamiltonian on the basis of $(|P_1\rangle, |P_2\rangle, |P_3\rangle, \ldots, |P_{12}\rangle, |C_1\rangle, |C_2\rangle, |C_3\rangle, \ldots, |C_{16}\rangle)^T$ is given by

$$H_K = \begin{bmatrix} H_K^P & H_K^{PG} \\ H_K^{PG\dagger} & H_K^G \end{bmatrix}, \qquad (7)$$

where $H_K^P$, $H_K^G$, and $H_K^{PG}$ denote the $K$-dependent Hamiltonians of phosphorene (a 12×12 matrix), graphene (a 16×16 matrix), and coupling between the layers (a 12×16 matrix), respectively, explicitly evaluated as

$$H_K^P = \begin{bmatrix}
0 & A_K^P & B_K^P & C_K^P & 0 & 0 & 0 & 0 & 0 & 0 & D_K^P & E_K^P \\
A_K^{P*} & 0 & C_K^{P*} & B_K^{P*} & 0 & 0 & 0 & 0 & 0 & 0 & F_K^P & G_K^P \\
B_K^{P*} & C_K^P & 0 & H_K^P & B_K^{P*} & C_K^P & 0 & 0 & 0 & 0 & 0 & 0 \\
C_K^{P*} & B_K^P & H_K^{P*} & 0 & C_K^{P*} & B_K^P & 0 & 0 & 0 & 0 & 0 & 0 \\
0 & 0 & B_K^P & C_K^P & 0 & A_K^P & B_K^P & C_K^P & 0 & 0 & 0 & 0 \\
0 & 0 & C_K^{P*} & B_K^{P*} & A_K^{P*} & 0 & C_K^{P*} & B_K^{P*} & 0 & 0 & 0 & 0 \\
0 & 0 & 0 & 0 & B_K^{P*} & C_K^P & 0 & H_K^P & B_K^{P*} & C_K^P & 0 & 0 \\
0 & 0 & 0 & 0 & C_K^{P*} & B_K^P & H_K^{P*} & 0 & C_K^{P*} & B_K^P & 0 & 0 \\
0 & 0 & 0 & 0 & 0 & 0 & B_K^P & C_K^P & 0 & A_K^P & B_K^P & C_K^P \\
0 & 0 & 0 & 0 & 0 & 0 & C_K^{P*} & B_K^{P*} & A_K^{P*} & 0 & C_K^{P*} & B_K^{P*} \\
D_K^{P*} & F_K^{P*} & 0 & 0 & 0 & 0 & 0 & 0 & B_K^{P*} & C_K^P & 0 & H_K^P \\
E_K^{P*} & G_K^{P*} & 0 & 0 & 0 & 0 & 0 & 0 & C_K^{P*} & B_K^P & H_K^{P*} & 0
\end{bmatrix}_{12\times 12}, \qquad (8)$$

$$H_K^G = \begin{bmatrix} 0 & A_K^G & t_1^C & B_K^G & t_2^C & 0 & 0 & 0 & 0 & 0 & 0 & 0 & C_K^G & 0 & D_K^G & E_K^G \\ A_K^{G*} & 0 & B_K^{G*} & t_1^C & 0 & t_2^C & 0 & 0 & 0 & 0 & 0 & 0 & 0 & C_K^G & F_K^G & D_K^G \\ t_1^C & B_K^G & 0 & t_1^C & t_1^C & B_K^G & t_2^C & 0 & 0 & 0 & 0 & 0 & 0 & 0 & C_K^G & 0 \\ B_K^{G*} & t_1^C & t_1^C & 0 & B_K^{G*} & t_1^C & 0 & t_2^C & 0 & 0 & 0 & 0 & 0 & 0 & 0 & C_K^G \\ t_2^C & 0 & t_1^C & B_K^G & 0 & A_K^G & t_1^C & B_K^G & t_2^C & 0 & 0 & 0 & 0 & 0 & 0 & 0 \\ 0 & t_2^C & B_K^{G*} & t_1^C & A_K^{G*} & 0 & B_K^{G*} & t_1^C & 0 & t_2^C & 0 & 0 & 0 & 0 & 0 & 0 \\ 0 & 0 & t_2^C & 0 & t_1^C & B_K^G & 0 & t_1^C & t_1^C & B_K^G & t_2^C & 0 & 0 & 0 & 0 & 0 \\ 0 & 0 & 0 & t_2^C & B_K^{G*} & t_1^C & t_1^C & 0 & B_K^{G*} & t_1^C & 0 & t_2^C & 0 & 0 & 0 & 0 \\ 0 & 0 & 0 & 0 & t_2^C & 0 & t_1^C & B_K^G & 0 & A_K^G & t_1^C & B_K^G & t_2^C & 0 & 0 & 0 \\ 0 & 0 & 0 & 0 & 0 & t_2^C & B_K^{G*} & t_1^C & A_K^{G*} & 0 & B_K^{G*} & t_1^C & 0 & t_2^C & 0 & 0 \\ 0 & 0 & 0 & 0 & 0 & 0 & t_2^C & 0 & t_1^C & B_K^G & 0 & t_1^C & t_1^C & B_K^G & t_2^C & 0 \\ 0 & 0 & 0 & 0 & 0 & 0 & 0 & t_2^C & B_K^{G*} & t_1^C & t_1^C & 0 & B_K^{G*} & t_1^C & 0 & t_2^C \\ C_K^{G*} & 0 & 0 & 0 & 0 & 0 & 0 & 0 & t_2^C & 0 & t_1^C & B_K^G & 0 & A_K^G & t_1^C & B_K^G \\ 0 & C_K^{G*} & 0 & 0 & 0 & 0 & 0 & 0 & 0 & t_2^C & B_K^{G*} & t_1^C & A_K^{G*} & 0 & B_K^{G*} & t_1^C \\ D_K^{G*} & F_K^{G*} & C_K^{G*} & 0 & 0 & 0 & 0 & 0 & 0 & 0 & t_2^C & 0 & t_1^C & B_K^G & 0 & t_1^C \\ E_K^{G*} & D_K^{G*} & 0 & C_K^{G*} & 0 & 0 & 0 & 0 & 0 & 0 & 0 & t_2^C & B_K^{G*} & t_1^C & t_1^C & 0 \end{bmatrix}_{16 \times 16}, \quad (9)$$

$$H_K^{PG} = \begin{bmatrix} t^{pc} & A_K^{PG} & t^{pc} & A_K^{PG} & t^{pc} & 0 & 0 & 0 & 0 & 0 & 0 & 0 & 0 & 0 & 0 & 0 \\ 0 & 0 & 0 & 0 & 0 & 0 & 0 & 0 & 0 & 0 & 0 & 0 & 0 & 0 & 0 & 0 \\ 0 & 0 & t^{pc} & t^{pc} & t^{pc} & t^{pc} & t^{pc} & t^{pc} & 0 & 0 & 0 & 0 & 0 & 0 & 0 & 0 \\ 0 & 0 & 0 & 0 & 0 & 0 & 0 & 0 & 0 & 0 & 0 & 0 & 0 & 0 & 0 & 0 \\ 0 & 0 & 0 & 0 & t^{pc} & A_K^{PG} & t^{pc} & A_K^{PG} & t^{pc} & A_K^{PG} & 0 & 0 & 0 & 0 & 0 & 0 \\ 0 & 0 & 0 & 0 & 0 & 0 & 0 & 0 & 0 & 0 & 0 & 0 & 0 & 0 & 0 & 0 \\ 0 & 0 & 0 & 0 & 0 & 0 & t^{pc} & t^{pc} & t^{pc} & t^{pc} & t^{pc} & t^{pc} & 0 & 0 & 0 & 0 \\ 0 & 0 & 0 & 0 & 0 & 0 & 0 & 0 & 0 & 0 & 0 & 0 & 0 & 0 & 0 & 0 \\ 0 & 0 & 0 & 0 & 0 & 0 & 0 & 0 & t^{pc} & 0 & t^{pc} & A_K^{PG} & t^{pc} & A_K^{PG} & t^{pc} & 0 \\ 0 & 0 & 0 & 0 & 0 & 0 & 0 & 0 & 0 & 0 & 0 & 0 & 0 & 0 & 0 & 0 \\ B_K^{PG} & 0 & 0 & 0 & 0 & 0 & 0 & 0 & 0 & 0 & 0 & 0 & t^{pc} & 0 & t^{pc} & t^{pc} \\ 0 & 0 & 0 & 0 & 0 & 0 & 0 & 0 & 0 & 0 & 0 & 0 & 0 & 0 & 0 & 0 \end{bmatrix}_{12 \times 16}, \quad (10)$$

whose elements are given by

$$A_K^P = t_5^p + t_2^p e^{-ik_a},$$

$$B_K^P = t_1^p + t_3^p e^{-ik_a},$$

$$C_K^P = 2t_4^p e^{-ik_a/2}\cos(k_a/2),$$

$$D_K^P = t_1^p e^{-ik_b} + t_3^p e^{-i(k_a+k_b)},$$

$$E_K^P = 2t_4^p e^{-ik_a/2} e^{-ik_b}\cos(k_a/2),$$

$$F_K^P = 2t_4^p e^{ik_a/2} e^{-ik_b}\cos(k_a/2),$$

$$G_K^P = t_1^{p*} e^{-ik_b} + t_3^p e^{i(k_a-k_b)},$$

$$H_K^P = t_2^p + t_5^p e^{-ik_a},$$

$$A_K^G = t_1^c e^{-ik_a},$$

$$B_K^G = 2t_2^c e^{-ik_a/2}\cos(k_a/2),$$

$$C_K^G = t_2^c e^{-ik_b},$$

$$D_K^G = t_1^c e^{-ik_b},$$

$$E_K^G = 2t_2^c e^{-ik_a/2} e^{-ik_b} \cos(k_a/2),$$

$$F_K^G = 2t_2^c e^{ik_a/2} e^{-ik_b} \cos(k_a/2),$$

$$A_K^{PG} = t^{pc} e^{-ik_a},$$

$$B_K^{PG} = t^{pc} e^{ik_b}, \tag{11}$$

where $k_a = \mathbf{K}.\mathbf{a}$ and $k_b = \mathbf{K}.\mathbf{b}$. Here $\mathbf{a} = a\hat{\mathbf{x}}$ and $\mathbf{b} = b\hat{\mathbf{y}}$, where $a = 4.43$ Å and $b = 9.81$ Å are the lattice constants. Diagonalizing the Hamiltonian matrix, we obtain the energy spectrum.

*Farhad Khoeini, khoeini@znu.ac.ir*